\documentstyle[12pt]{article}

\catcode`\@=11 
\newcount\@tempcntc
\def\@citex[#1]#2{\if@filesw\immediate\write\@auxout{\string\citation{#2}}\fi
  \@tempcnta\z@\@tempcntb\m@ne\def\@citea{}\@cite{\@for\@citeb:=#2\do
    {\@ifundefined
       {b@\@citeb}{\@citeo\@tempcntb\m@ne\@citea\def\@citea{,}{\bf ?}\@warning
       {Citation `\@citeb' on page \thepage \space undefined}}%
    {\setbox\z@\hbox{\global\@tempcntc0\csname b@\@citeb\endcsname\relax}%
     \ifnum\@tempcntc=\z@ \@citeo\@tempcntb\m@ne
       \@citea\def\@citea{,}\hbox{\csname b@\@citeb\endcsname}%
     \else
      \advance\@tempcntb\@ne
      \ifnum\@tempcntb=\@tempcntc
      \else\advance\@tempcntb\m@ne\@citeo
      \@tempcnta\@tempcntc\@tempcntb\@tempcntc\fi\fi}}\@citeo}{#1}}
\def\@citeo{\ifnum\@tempcnta>\@tempcntb\else\@citea\def\@citea{,}%
  \ifnum\@tempcnta=\@tempcntb\the\@tempcnta\else
   {\advance\@tempcnta\@ne\ifnum\@tempcnta=\@tempcntb \else \def\@citea{--}\fi
    \advance\@tempcnta\m@ne\the\@tempcnta\@citea\the\@tempcntb}\fi\fi}
\catcode`\@=12 
\newcommand{\smallHT}{{\scriptscriptstyle HT}}
\newcommand{\eqref}[1]{(\ref{#1})}   
\newcommand{\naive}{na\"{\i}ve}

\newcommand{\Du}{\hbox{$\Delta u$}}
\newcommand{\Dd}{\hbox{$\Delta d$}}
\newcommand{\Ds}{\hbox{$\Delta s$}}
\newcommand{\DSigma}{\hbox{$\Delta \Sigma$}}

\newcommand{\gIp}{g_1^p(x,Q^2)}

\newcommand{\GIp}{\Gamma_1^p(Q^2)}
\newcommand{\GIn}{\Gamma_1^n(Q^2)}

\newcommand{\pr}{\paragraph{}}
\newcommand{\beq}{\begin{equation}}
\newcommand{\eeq}{\end{equation}}
\newcommand{\bea}{\begin{eqnarray}}

\newcommand{\eea}{\end{eqnarray}}

\newcommand{\OO}{{\cal O}}
\newcommand{\Oas}[1]{\hbox{${\cal O}(
\left({\alpha_s/\pi})^{#1}\right)$}}
\newcommand{\OI}{\hbox{${\cal O}({\alpha_s/\pi})$}}
\newcommand{\OII}{\Oas{2}}
\newcommand{\OIII}{\Oas{3}}
\newcommand{\OIV}{\Oas{4}}
\newcommand{\MSbar}{{\overline{MS}}}
\newcommand{\LamMSbar}{\Lambda_{\MSbar}}
\newcommand{\asMZ}{\hbox{$\alpha_s(M_Z^2)
\bigl\vert\bigr.\lower.5em\hbox{$_\MSbar$}$}\ }
\def\undertext#1{{\em #1}}

\begin{document}

\begin{titlepage}
\begin{flushright}
CERN-TH-7324/94\\
TAUP-2178-94\\
hep-ph/9407287
\end{flushright}
\begin{centering}
\vspace{.1in}
{\large {\bf Determination of $\alpha_s$ and the Nucleon\\
Spin Decomposition using Recent Polarized Structure Function Data}}\\
\vspace{.4in}
{\bf John Ellis}\\
\vspace{.05in}
Theory Division, CERN, CH-1211, Geneva 23, Switzerland. \\
e-mail: johne@cernvm.cern.ch \\
\vspace{0.5cm}
and \\
\vspace{0.5cm}

\vspace{.05in}
{\bf Marek Karliner}\\

\vspace{.05in}
School of Physics and Astronomy
\\ Raymond and Beverly Sackler Faculty of Exact Sciences
\\ Tel-Aviv University, 69978 Tel-Aviv, Israel
\\ e-mail: marek@vm.tau.ac.il
\\
\vspace{.03in}
\vspace{.1in}
{\bf Abstract} \\
\vspace{.05in}
\end{centering}
{\small

New data on polarized $\mu{-}p$ and $e{-}p$
scattering permit a first determination of $\alpha_s$
using the Bjorken sum rule, as well as higher
precision in determining the nucleon spin
decomposition. Using perturbative QCD calculations
to \OIV\ for the non-singlet combination of
structure functions, we find
$\alpha_s(2.5\,\, \hbox{GeV}^2) = 0.375^{+0.062}_{-0.081}\,\,$,
corresponding to $\alpha_s(M_Z^2) =0.122^{+0.005}_{-0.009}\,\,$,
and using calculations to \OIII\ for the singlet combination
we find
$\Du = 0.83  \pm 0.03$,
$\Dd =-0.43  \pm 0.03$,
$\Ds =-0.10  \pm 0.03$,
$\Delta \Sigma \equiv \Du + \Dd + \Ds =  0.31 \pm 0.07$,
at a renormalization scale $Q^2 = 10\ \hbox{GeV}^2$.
Perturbative QCD corrections play an essential role
in reconciling the interpretations of
data taken using different targets.
We discuss higher-twist uncertainties in these determinations.
The $\Delta q$ determinations are used to update predictions for
the couplings of massive Cold Dark Matter particles and axions
to nucleons.
}
\paragraph{}
\par
\vspace{0.7in}
\begin{flushleft}
CERN-TH-7324/94\\
July 1994 \\
\end{flushleft}

\end{titlepage}
\newpage

When Bjorken first derived his sum rule for polarized deep-inelastic
scattering \cite{BjI},
he doubted whether it could be tested experimentally.
Several years later, however, he changed his mind \cite{BjII}
and an experimental programme
on polarized electron-proton scattering was launched at
SLAC \cite{oldSLACa,oldSLACb,oldSLACc}
In view of this, a sum rule for the polarized proton
and neutron
structure function $g_1^p$ was proposed \cite{EJ},
based on the dynamical
hypothesis that strange quarks and antiquarks in a polarized
nucleon would have no net polarization. However, this sum rule
was not at all rigorous, in contrast to the Bjorken sum rule
which is an inescapable prediction of
QCD \cite{Kodaira,KodairaAnomDim}.
The first round of
polarized $e$-$p$ experiments that were consistent, within their
stated errors, with the hypothesis of no net strange polarization.
However, no measurements were made on polarized $e$-$n$
scattering, so the crucial Bjorken sum rule went untested.

Several years later, following the advent of naturally-polarized
high-energy and \hbox{-intensity} muon beams at CERN, the EMC made a
second measurement of $g_1^p$ \cite{EMCPL,EMCNP}.
  This was more precise than the
earlier SLAC measurements, and extended to lower values of
$x\equiv \vert q^2 \vert/2 m_p \nu$. It indicated a disagreement
with the polarized-proton sum rule, corresponding to a non-zero
and negative net contribution \Ds\ of strange quarks and antiquarks
to the spin of the proton, and a small total contribution of the
light quarks: $\Du + \Dd + \Ds \ll 1$\quad
\cite{SmallSpinI,SmallSpinII}.
This result was surprising
from the point of view of the most \naive\ formulation of the
constituent quark model, which would suggest $\Du + \Dd \simeq 1$, and
$\Ds = 0$, and even compared with more sophisticated versions
adapted to fit measurements of $g_A = \Du - \Dd$ and the
hyperon axial-current matrix elements' $F/D$ ratio, which suggested
$\Du + \Dd \simeq 0.6$ and $\Ds \simeq 0$.

This surprise indicated that our theoretical understanding of
non-perturbative QCD was incomplete, and stimulated attempts
to remedy this defect. For example, it was pointed out
\cite{BEK} that the
EMC result could be understood qualitatively within $SU(3)_f$
topological models of the nucleon, which describe light quarks via
collective fields, incorporate the global symmetries
of non-perturbative QCD, and predict $\Du + \Dd + \Ds \ll 1$.
Alternatively, it was suggested
\cite{DeltagI,DeltagII,DeltagIII}
that the $U(1)$ axial anomaly
might play a key r\^ole, which would modify the \naive\ quark
model predictions.
It was even suggested that the ``sacred"
Bjorken sum rule might be violated \cite{BJviolation}.

In view of the interest in checking the EMC polarized
$\mu$-$p$ results and extending them to test the Bjorken sum rule,
extensive experimental programmes are under way at CERN, SLAC and
DESY. An important round of results on polarized $\mu$-$D$
scattering from CERN \cite{SMCd}
and polarized $^3$He scattering from SLAC \cite{E142} were
published in 1993, permitting for the first time an
experimental test of the Bjorken sum rule. The CERN and SLAC
data agreed within their errors on the extracted value of
$g_1^n$
and confirmed the validity of the Bjorken sum rule
with a precision of about 12\% \cite{EKBjSR},
once the then-available
higher-order perturbative QCD corrections \cite{BJcorr}
were taken into account and allowance made for higher-twist
corrections \cite{HigherTwist,BBK}.
These data also indicated $\Ds < 0$ and
$\Du + \Dd +\Ds \ll 1$.

Recently, two new sets of data with polarized proton targets have
been made available \cite{SMCp,E143}.
 In addition, order $(\alpha_s/\pi)^2$
corrections to the singlet part of structure functions
$g_1^p$ and $g_1^n$ have recently been calculated \cite{Larin}
and estimates
made of higher-order perturbative QCD corrections to
the Bjorken and singlet
sum rules \cite{EstimCorr,KataevSinglet}.
Therefore, now is an appropriate moment to reassess the
precision with which the Bjorken sum rule has been tested,
and to extract new estimates of \Du, \Dd\ and \Ds.

We find that the Bjorken sum rule is now verified with a
precision of
10\%, once all available perturbative corrections are
taken into account. As an exercise, we determine $\alpha_s$
for the first time from the polarization sum rule data,
finding
$\alpha_s(2.5\,\, \hbox{GeV}^2) = 0.375^{+0.062}_{-0.081}\,\,$,
corresponding to $\alpha_s(M_Z^2) =0.122^{+0.005}_{-0.009}\,\,$.
On the other hand, we find
a consistent pattern of violation of the
separate sum rules for proton and neutron targets \cite{EJ},
indicating that $\Ds < 0$. Encouraged by the consistency,
we extract values of $\Du + \Dd + \Ds$ from the
different data sets. These are superficially different
if the data are analyzed in the \naive\ parton model,
i.e. neglecting perturbative QCD corrections. However,
the agreement between different data sets improves
systematically as each higher order of perturbative
QCD correction is included.
Including the \OIII\ calculation \cite{BJcorr}, together with
estimates of \OIV\ effects
in the non-singlet channel \cite{EstimCorr}
and
the \OII\ calculation \cite{Larin},
together with estimates of \OIII\
effects in the singlet channel \cite{KataevSinglet},
we find
\beq
        \Du = 0.83  \pm 0.03,
 \quad \Dd =-0.43  \pm 0.03,
 \quad \Ds =-0.10  \pm 0.03,
 \quad \Delta \Sigma = 0.31 \pm 0.07
\label{finalDqs}
\eeq
at a renormalization scale $Q^2 = 10\ \hbox{GeV}^2$,
with a global $\chi^2=1.3\ $ for $5$ degrees of freedom.
As an example of relevance of these results, we present
at the end of this paper an analysis of the implications of these
determination for dark matter couplings to matter, including
the elastic scattering of supersymmetric relics and
axion couplings to nucleons.

Up to now, the prevailing attitude has been to use
polarized structure function data to test the Bjorken
sum rule, using the most complete theoretical calculations
of corrections and a world-average value of $\alpha_s$. The
conclusion last year was \cite{EKBjSR}
that the EMC/SMC and SLAC E142
data together verified the Bjorken sum rule to within
the available precision of 12\%.
Here we take a different attitude, more akin to that adopted
with regard to the Gross-Llewellyn Smith sum rule \cite{GLLSsr}.
There is no convincing theoretical evidence against the validity
of the Bjorken sum rule, which is a solid prediction of QCD.
Therefore, its validity can be assumed. Instead,
one can use the new polarized structure function
data to extract a value of $\alpha_s(Q^2)$, whose
consistency with other measurements is an {\em a posteriori} check
on the validity of the Bjorken sum rule.

As is well known, data at low $Q^2$ are particularly
sensitive to $\LamMSbar$ and hence at a premium in determining
\asMZ. As is illustrated by
$\tau$ decays, even a relatively imprecise determination
of $\alpha_s(\hbox{low}\ Q^2)_{\MSbar}$
(provided higher-order perturbative QCD and nonperturbative
uncertainties are understood) extrapolates to a
relatively precise determination of \asMZ\ \cite{alphasALEPH}.
The SLAC E142 $^3$He (i.e. $n$) data have
$\langle Q^2 \rangle \simeq 2$ GeV$^2$,
and the E143 $p$ data have
$\langle Q^2 \rangle \simeq 3$ GeV$^2$,
and are hence particularly well placed to exploit this
lever arm on $\LamMSbar$. Since they are at higher $Q^2$,
the new SMC $p$ data are less useful in this respect, though they
do provide important information about $\gIp$
at small $x$, and enter into the determination
of the $\Delta q$, as we discuss later.

We use the value of $\Gamma_1^n(Q^2{=}2\ \hbox{GeV}^2)=
-0.028 \pm 0.006 \pm 0.009$ given in ref.~\cite{EKBjSR}
and the value  $\Gamma_1^p(Q^2{=}3\ \hbox{GeV}^2)=
0.133 \pm 0.004 \pm 0.012$ given in ref.~\cite{E143}.
We emphasize that these estimates are based on polarization
asymmetry $A_1(x,Q^2)$ data taken at average values of $Q^2$
that depend on the bin in $x$, which have been converted
into values of $g_1^p(x,\langle Q^2 \rangle)$
by assuming that the $Q^2$ dependence of $A_1(x,Q^2)$
is insignificant  \cite{EKBjSR,Petratos}
and using standard parametrizations of
$F_2(x,Q^2)$ \cite{NMC} and $R(x,Q^2)$ \cite{Whitlow}
to estimate
\beq
g_1(x,Q^2) =
{A_1(x)F_2(x,Q^2)\over 2x [1 + R(x,Q^2)]}
\label{gIapprox}
\eeq
No data set indicates a significant $Q^2$ dependence of
$A_1(x,Q^2)$, and perturbative QCD calculations
\cite{ZvN,ANR,BBS,GS} lead one to expect that the
$Q^2$ dependence can be neglected at the level of
precision required here.
The estimates of $g_1^{p,n}(x,Q^2)$ also require
assumptions on $g_2^p(x,Q^2)$ that are borne out
by the latest experimental bounds \cite{SMCp},
and consistent with the latest theoretical
calculations \cite{ALNR}.
Based on the above numbers, we use in our subsequent
analysis
\beq
\Gamma_1^p - \Gamma_1^n \vert_{Q^2=2.5\ \hbox{\scriptsize GeV}^2}
=0.161 \pm 0.007 \pm 0.015
\label{Gammapn}
\eeq
for the Bjorken integral.

    The small-$x$ behaviours of $g_1^{p,n}$ have recently been
discussed \cite{BassLandshoff}
in the context of a non-perturbative model
of the Pomeron \cite{NonPertP},
which suggests that ${g_1^{p,n} \simeq
\log (1/x)}$ at small $x$,
rather than $\sim x^{\alpha}$, $-0.5 \le \alpha \le 0$
\cite{Heimann,EK}.
Although such an effect would alter the
estimates of $\Gamma_1^{p,n}$ that we use, it would cancel out in
the difference that appears in the Bjorken sum rule, since this
effect would be due to two-gluon exchange. Hence the estimate in
equation \eqref{Gammapn} would be unaltered.

Perturbative QCD corrections to the Bjorken sum rule
have been calculated up to \OIII\ \cite{BJcorr}
and an estimate was made of the \OIV\ coefficient
\cite{EstimCorr}:
\bea
\Gamma_1^p(Q^2) - \Gamma_1^n(Q^2) =
{1\over6}\vert g_A \vert
\left[ 1 - \left(\frac{\alpha_s(Q^2)}{\pi}\right)
    -3.5833 \left(\frac{\alpha_s(Q^2)}{\pi}\right)^2
\right.\nonumber\\
\label{BJpredIV}\\
\left.
-20.2153 \left(\frac{\alpha_s(Q^2)}{\pi}\right)^3
-\OO(130) \left(\frac{\alpha_s(Q^2)}{\pi}\right)^4
+ \dots \right]\nonumber
\eea
in the $\MSbar$ prescription for $N_f=3$ flavours, as
appropriate to $Q^2=2.5$ GeV$^2$. The value of
$\alpha_s(2.5\ \hbox{GeV}^2)$ extracted by comparing
\eqref{Gammapn} and \eqref{BJpredIV} depends on the
order in QCD perturbation theory which is used. As we
see in fig.~1, the extracted value decreases as one
progresses from \OI\ to higher orders, but
shows good signs of stabilizing in \OIV. We infer from
this analysis a value
\beq
\alpha_s(2.5\,\, \hbox{GeV}^2)\vert_{\MSbar, N_f=3}
 = 0.375^{+0.062}_{-0.081}
\label{aslowQeq}
\eeq
which corresponds to
\beq
\alpha_s(M_Z^2)\vert_{\MSbar,N_f=5}
 =0.122^{+0.005}_{-0.009}
\label{asmZeq}
\eeq
Note that in evaluating \eqref{asmZeq}\ we have used the
3-loop renormalization group equations \cite{ThreeLoopBeta},
and matched
values of $\alpha_s$ at the flavour thresholds
$Q=m_c, m_b$, as is appropriate
 in the $\MSbar$ scheme.
The relevant prescription is given in ref.~\cite{Bernreuther}:
${\alpha_s}_{-}(m_Q^2)=
 {\alpha_s}_{+}(m_Q^2)
\bigl[1+ a\,({\alpha_s}_{+}(m_Q^2)/\pi)
  + b\,({\alpha_s}_{+}(m_Q^2)/\pi)^2\bigr]$,
  where $a=0$, $b=7/72$.
This means that the two-loop
 ${\alpha_s}_{-}(m_Q^2)={\alpha_s}_{+}(m_Q^2)$,
but that the three-loop
 ${\alpha_s}_{-}(m_Q^2)$ and ${\alpha_s}_{+}(m_Q^2)$
are slightly different at $m_Q$.
Figure 3 compares this determination of $\alpha_s$ with
others, as compiled in \cite{Bethke}.

    We have not yet included higher-twist effects
\cite{HigherTwist}
in this analysis,
for two main reasons. One is that their coefficients appear to be
rather smaller than originally thought \cite{BBK}, and
also because the available estimates differ considerably from
one another
\cite{CloseRoberts,Ji,BI,ESM,RossRoberts,ZZ,MSS}.
For our purposes it is sufficient to take
a rough estimate,
\beq
\delta_\smallHT\, ( \Gamma_1^p - \Gamma_1^n ) \equiv
{c_{\smallHT} \over Q^2} \simeq
{( -0.02 \pm 0.01) \ \hbox{GeV}^2 \over  Q^2 }
\label{DeltaHT}
\eeq
Another reason is that it may be double counting \cite{Mueller}
to include this as
well as the higher-order perturbative QCD contributions that have
now been calculated and estimated. Including the estimate
\eqref{DeltaHT} in
a fit to the data described above, we find
\beq
\alpha_s(M_Z^2)\vert_{\MSbar,N_f=5,\smallHT}
= 0.118^{+0.007}_{-0.014}
\label{asmZhteq}
\eeq
The difference between this and the value in equation
\eqref{asmZeq} is
indicative of the theoretical systematic error.

This Bjorken sum rule determination of $\alpha_s$ is
encouragingly precise, and quite consistent with the
other determinations shown in fig.~3, though not yet
competitive with the market leaders, This consistency
means that the Bjorken sum rule is verified within
10\% precision by the SLAC E142 and E143 data alone, as well
as being verified to within 12\% precision by the SLAC
E142 and EMC/SMC data \cite{EKBjSR}.
The higher precision expected from future SLAC data might
enable this determination of $\alpha_s$ to become truly
competitive.
A good understanding of the radiative corrections will be
needed at this level \cite{Krasny}.
The optimal evaluation of the Bjorken integral will also
require data from low $x$, where the SMC data are available
and will continue to provide crucial input. Also, the very
high precision data from HERMES \cite{HERMES}
at moderate $x$ will be
very useful. The Bjorken sum rule is making the
transition from a test of QCD to a tool for
evaluating $\alpha_s$ within QCD.

Having verified that the Bjorken sum rule is well satisfied
by the latest data, we consider the sum rules for proton and neutron
targets separately. Their theoretical prediction requires an extra
assumption on the singlet axial current matrix elements
$\Delta\Sigma(Q^2)$ or, equivalently, the nucleon matrix element \Ds\ of
the $\bar{s} \gamma_\mu\gamma_5 s$ current. Alternatively, one can
regard measurements of
$\Gamma_1^{p(n)} \equiv \int_0^1 d x \, g_1^{p(n)}(x,Q^2)$
as determinations of $\Delta\Sigma(Q^2)$, or
equivalently \Ds, if one assumes the validity of the Bjorken sum rule.
This is the attitude taken here.
The fact that $\Gamma_1^p-\Gamma_1^n$ is in good agreement
with the Bjorken sum rule means that the different targets
yield consistent values of $\DSigma$, as we now show.

We use in our determinations of $\Delta\Sigma(Q^2)$ and \Ds\ the
calculated \OIII\ corrections to the Bjorken sum rule and the
estimate \cite{EstimCorr}
of the \OIV\ corrections used above.
We also use the recent calculation of the \OII\ correction
\cite{Larin} to the
singlet part of the sum rule and a recent estimate \cite{KataevSinglet}
of the
\OIII\ correction to it. If one chooses the renormalization
scale $\mu = Q$, the proton and neutron sum rules including these
corrections can be written as
\bea
\int_{0}^{1} d x g_1^{p(n)}(x,Q^2)=
\left(\pm \frac{1}{12}|g_A|+\frac{1}{36}a_8\right)\times
\phantom{aaaaaaaaaaaaaaaaaaa}
\nonumber\\
\times\left[ 1 {-} \left(\frac{\alpha_s(Q^2)}{\pi}\right)
\kern-0.2em
{-}3.5833 \left(\frac{\alpha_s(Q^2)}{\pi}\right)^{\kern-0.3em2}
\kern-0.3em
{-}20.2153 \left(\frac{\alpha_s(Q^2)}{\pi}\right)^{\kern-0.3em3 }
\kern-0.3em
{-}\OO(130)\left(\frac{\alpha_s(Q^2)}{\pi}\right)^{\kern-0.3em4}
\kern-0.3em
{+}\dots\,\right]
\phantom{}\kern-0.5cm 
\nonumber
\\
+\left[1 - \left(\frac{\alpha_s(Q^2)}{\pi}\right)
-1.0959 \left(\frac{\alpha_s(Q^2)}{\pi}\right)^2
-{\cal O}(6)\left(\frac{\alpha_s(Q^2)}{\pi}\right)^3
+\dots\,\,\right]
 \frac{1}{9} \Delta\Sigma(Q^2)\,.
\nonumber\\
\label{nucleonSR}
\eea
It is worth emphasizing that the
$Q^2$ dependence due to QCD in the singlet channel has
two sources, namely the anomalous dimension of the
singlet axial current, as well as the coefficient
function. However, the $Q^2$ dependence of $\DSigma(Q^2)$
is not large from the $Q^2$ range of the present data
upwards.
Note that we have not included any higher-twist contributions, but
will comment later on their possible effects.

It is interesting to note that, because of their different
$Q^2$-dependences, it would be possible in principle to
separate the $SU(3)$ octet and singlet contributions
$(a_8,\,\Delta\Sigma)$ to either $\GIp$ or $\GIn$, and avoid in
this way the use of $SU(3)$ \cite{LipkinWeak}
relations to estimate $a_8$ on the
basis of hyperon $\beta$-decay data.
Writing the square-bracket perturbative QCD correction factors
on the first and last lines of equation \eqref{nucleonSR}
as $f(\alpha_s)$ and $h(\alpha_s)$, respectively, for the
octet and singlet contributions, one can write
\bea
\nonumber\\
{f(\alpha_s)\over 36} a_8 \quad + \quad {h(\alpha_s)\over 9 }
\DSigma(Q^2)
\quad  = \quad
\Gamma_1^p(Q^2) \quad - \quad {f(\alpha_s)\over 12} g_A
\nonumber\\
\nonumber\\
{f(\alpha_s)\over 36} a_8 \quad + \quad {h(\alpha_s)\over 9 }
\DSigma(Q^2)
\quad = \quad
\Gamma_1^n(Q^2) \quad + \quad {f(\alpha_s)\over 12} g_A
\label{GpGnGdA}\\
\nonumber\\
{f(\alpha_s)\over 36} a_8 \quad + \quad {h(\alpha_s)\over 9 }
\DSigma(Q^2)
\quad = \quad
\Gamma_1^d(Q^2) \phantom{\quad + \quad {f(\alpha_s)\over 12} g_Aa}
\nonumber\\
\nonumber
\eea
where the right-hand sides are combinations of measurable and
theoretically-known quantities.
(Strictly speaking, $\Gamma_1^d$ in eq.~\eqref{GpGnGdA}
is $(\Gamma_1^p+\Gamma_1^n)/2$,
i.e. it includes nuclear corrections.)
In leading order,
$f(\alpha_s)=h(\alpha_s)=1-(\alpha_s/\pi)$, and the
different equations \eqref{GpGnGdA} are not independent.
At NLO, however, $f(\alpha_s) \neq h(\alpha_s)$, and then
\undertext{each} of the equations \eqref{GpGnGdA} allows in
principle an independent determination of $\DSigma$ and
$a_8$, provided data with sufficiently high precision are
available at different values of $Q^2$. One can also combine
data from experiments with different targets. The
current precision of the
world data does not permit a meaningful separation of $a_8$
and $\DSigma$ using this technique.

Using
the known variation of $\alpha_s(Q^2)$ and
assuming $\DSigma \sim 0.3$ in
eq.~\eqref{GpGnGdA},
one can estimate the experimental
precision required to make this approach practical.
For a range of $Q^2$ between $Q^2{=}2$ GeV$^2$
and $Q^2{=}15$ GeV$^2$
we estimate that the right-hand side of
eq.~\eqref{GpGnGdA} would have to be known with
a precision much better than $\pm0.002$, including both
statistical and systematic errors.
Currently E143 quotes
\cite{E143}
error values of
$\pm0.004$ (stat.) and
$\pm 0.012$ (syst.), so the required
improvement in precision would have to be very substantial,
especially in the systematic error, unless for some reason
the systematic error is approximately independent of
$Q^2$, in which case it would largely cancel out
when comparing the r.h.s. of eq.~\eqref{GpGnGdA}
for the various values of $Q^2$.
Pending future improvements in the experimental precision,
for the rest of this paper we assume the usual
$SU(3)$ value $a_8 = 0.601 \pm 0.038$ \cite{FoverD},
and we return now to the discussion of the existing data.

In order to combine all the data available from different
targets, we re-express the measurement on Deuteron and $^3$He
targets in terms of the corresponding values of $\GIp$,
assuming the Bjorken sum rule and incorporating all the non-singlet
perturbative corrections $f(\alpha_s)$ on the first line of
equation \eqref{nucleonSR}. The resulting values of $\GIp$
inferred from the E142  $^3$He and SMC D data are shown together
with the EMC, SMC and preliminary E143  $p$ data in fig.~4. We
see \undertext{no signs of convergence} towards the
{\naive}ly-suggested \cite{EJ} value holding if $\Ds = 0$.
This situation contrasts with what has been found previously
with the Gross - Llewelyn Smith and Bjorken sum rules, namely
\undertext{good agreement} once the available perturbative QCD
corrections are included. Also shown in fig.~4, to guide the
eye, is the prediction for $\GIp$ that would be obtained
if $\Ds = -0.10 \pm 0.04$ which is \undertext{highly consistent}
with the data. The inclusion of all available higher-order
perturbative QCD corrections is important for this consistent
picture to emerge. We plot in fig.~5 the values of $\DSigma$
that would be extracted from each experiment if one restricted
one's analysis to include only low orders of perturbative QCD.
The values of $\DSigma$ extracted using the \naive\ parton model,
i.e. \Oas{0}, are in poor agreement, particularly the E142
neutron measurement done on $^3$He. However, the agreement
improves significantly as one proceeds to \OI, \OII and
\OIII. (This last analysis includes the estimate of the
\OIV\ corrections to the Bjorken sum rule \cite{EstimCorr}
and of the \OIII\ correction \cite{KataevSinglet}
to the singlet sum rules.)
The overall $\chi^2$ of the global fit decreases systematically
as each order of perturbative QCD is included,
$\chi^2 = 12$ (\naive) $\rightarrow$ 4.2 (\OI)
$\rightarrow$ 2.4 (\OII) $\rightarrow$ 1.6 (\OIII)
$\rightarrow$ 1.3 (\OIV).
The fact that the $g_1^n$ determination of $\DSigma$
falls in increasing orders of perturbation theory,
whilst the determination from $g_1^p$
rises, is easily understood from the fact that
the higher order perturbative QCD corrections to the
Bjorken sum rule are much larger than those of the
singlet combinations of structure functions.
The main perturbative correction to the non-singlet part
comes from the isovector term $\pm g_A$ which reverses
its sign between the neutron and the proton. In the deuteron
this term is absent and the effect of the remaining
$a_8$ term is small, and therefore
 the $g_1^D$ determination does not
change much in increasing orders of perturbation theory.

We conclude that a consistent overall picture
of $\DSigma$ and $\Ds$ is emerging, which leads to the
values quoted in equation \eqref{finalDqs} at the
beginning of this paper.

    The overall consistency of the different determinations of
$\DSigma$ and ${\Delta s}$ that we find is not affected by the
possible $\log(1/x)$ behaviour \cite{BassLandshoff}
of the $g_1^{p,n}$ that
we discussed earlier, since this consistency is simply a
consequence of the data and the assumed correctness of the Bjorken
sum rule. However, such behaviour would alter the determinations
of the individual ${\Delta q}$ quoted in equation \eqref{finalDqs}.
As pointed out in \cite{CloseRobertsG1},
this could shift the experimental values of the
integrals $\Gamma_1^{p,n}$ and hence the inferred values of the
${\Delta q}$ by about one standard deviation towards a higher
value of $\DSigma$ and a lower value of $\Ds$. As also
pointed out in \cite{CloseRobertsG1},
these shifts could be much larger if the
small-$x$ behaviours of $g_1^{p,n}$ were even more singular, but
this possibility does not seem to be very well motivated theoretically.
For the reasons discussed earlier in connection with the Bjorken
sum rule, we have not yet included estimates of singlet
higher-twist effects
\beq
\delta_\smallHT\, ( \Gamma_1^p + \Gamma_1^n )_{singlet} \simeq
{( -0.02 \pm 0.01) \ \hbox{GeV}^2 \over  Q^2 }
\label{dHTsinglet}
\eeq
in the extraction of the ${\Delta q}$. If we include them in the
global fit, we find
\beq
       \Du = 0.85  \pm 0.03,
 \quad \Dd =-0.41  \pm 0.03,
 \quad \Ds =-0.08  \pm 0.03,
 \quad \Delta \Sigma = 0.37 \pm 0.07
\label{DqwithHT}
\eeq
at ${Q^2=10\ \hbox{GeV}^2}$, and
we regard the differences between these and the values in equation
\eqref{finalDqs}
as indicative of the possible theoretical systematic errors.

We cannot resist commenting that the value of $\DSigma$
that we obtain is much closer to the light current-quark
model value of zero in the large-$N_c$ and chiral
limit \cite{BEK}
than it is to the most \naive\ quark model of unity, and that
finite-$N_c$ \cite{Ryzak} and $m_s \neq 0$ corrections
\cite{BEK},\cite{SkyrmeNow} were estimated to be capable
of altering $\DSigma$ by about 0.3\ .
This picture may be reconciled with otherwise
highly-successful models based on constituent quarks, if the
latter are regarded as effective constructs that
contain non-trivial internal chiral and gluonic
structure \cite{MG,Kaplan,Fritzsch,EFHK,staticQ,Weise}.

    Finally, as an application of these results, we discuss their
implications for the couplings to nucleons of certain candidates
for non-baryonic Dark Matter. We consider first the lightest
supersymmetric particle (LSP), which we assume to be a neutralino,
i.e., some combination of neutral, non-strongly-interacting spin-1/2
partners of the $SU(2)$ gauge boson $W^0$,
the $U(1)$ gauge boson $B$,
and the neutral Higgs boson $H^0_{1,2}$ in the minimal supersymmetric
extension of the Standard Model (MSSM). Since the neutralino has
spin $1/2$, its couplings to nucleons include a spin-dependent
part that is related to the contributions ${\Delta q}$
of the different
quark species to the nucleon spin. Experiments at LEP and elsewhere
constrain the parameters of the MSSM in such a way that the LSP
cannot be an approximately pure photino or Higgsino. However, it
could be an approximately pure $U(1)$ gaugino
${\tilde B}$, in which case its
spin-dependent coupling to a proton is proportional to
\beq
a_p = 17/36 \Delta u + 5/36 ( \Delta d + \Delta s )
= ( 6 g_A + 2 a_8 + 9\DSigma )/36
\label{apDef}
\eeq
and the analogous coupling $a_n$ to the neutron is given by a
similar formula with ${\Delta u}$ and ${\Delta d}$ interchanged,
in the limit of a small momentum transfer from the LSP to the
proton and $m_{LSP} \ll m_{\tilde q}$. The couplings $a_{p,n}$
should be evaluated
using the value of the ${\DSigma}$ evolved to a renormalization
scale of order $m_{\tilde q}$,
which we take for illustration
to be 500 GeV. Using the values (and errors) of the ${\Delta q}$
given in equation \eqref{finalDqs}, we find
\bea
a_p =  \phantom{{-}}0.32 \pm 0.017   \nonumber\\
\label{apanValues} \\
a_n = {-}0.10 \pm 0.017  \nonumber
\eea
We note that the error on $a_p$ is only about 5\%, whilst the
error on $a_n$ is about 17\%. This implies that one can estimate
relatively reliably the spin-dependent coupling of the LSP to
odd-even nuclei such as $^{39}$K or $^{93}$Nb,
whose spins are carried essentially by protons, whereas
the uncertainty is somewhat larger for even-odd nuclei
such as $^{73}$Ge or $^{29}$Si, whose spins
are carried essentially by neutrons.
Predictions based on the \naive\
quark model \cite{KKR}
for the spin content of the nucleon are very
misleading, particularly for the neutron.
\pr
Next, we turn to the couplings of the axion to nucleons, which are
given by
\bea
C_{ap} = 2 \,[{-} 2.76\, \Du - 1.13\, \Dd + 0.89\, \Ds
-\cos 2 \beta \, (\Du - \Dd - \Ds) ],
\nonumber\\
\label{CapCan}\\
C_{an} = 2\, [{-} 2.76\, \Dd - 1.13\, \Du + 0.89\, \Ds
-\cos 2 \beta \, (\Dd - \Du - \Ds) ]
\phantom{,}
\nonumber
\eea
for the proton and neutron respectively. In this case, the
low-renormalization scale values of the ${\Delta q}$ are the
relevant ones, and we evaluate them at the scale $Q = 1$ GeV,
which might be appropriate for the core of a neutron star. We find
\bea
C_{ap} = ({-}3.9 \pm 0.4) - (2.68 \pm 0.06)
\cos 2 \beta
\nonumber\\
\label{Vanalogue}\\
C_{an} = (0.19 \pm 0.4) + (2.35 \pm 0.06)
\cos 2 \beta
\phantom{,}
\nonumber
\eea
which can be compared with the \naive\ quark model values given in
equation (4) of \cite{Mayle}. We see that the axion-proton
coupling is relatively well determined, except for large values of
$\tan \beta$, whilst the axion-neutron coupling
is more sensitive at intermediate values of $\tan \beta$,
to the experimental errors in the determination
of the ${\Delta q}$. As was discussed in \cite{Mayle}
and references therein, the
dominant axion emission process from the core of a neutron star
is likely to be axion bremsstrahlung in nucleon-nucleon collisions,
which is sensitive to a combination of $C_{ap}$ and $C_{an}$. The
relative weights of neutron-neutron, neutron-proton and proton-proton
bremsstrahlung processes are given roughly by
\beq
{C_{an}}^2 + 0.83 (C_{an} + C_{ap})^2 + 0.47 {C_{ap}}^2
\label{Csquares}
\eeq
One
tries to bound the axion decay constant $f_a$ by constraining the
rate of axion emission from the core of the neutron star. Using
the numbers given in equation \eqref{Vanalogue}, we find a
sensitivity
\beq
\Delta f_a / f_a = 30\% \ \hbox{to}\ 50\%
\label{fSensitivity}
\eeq
This source of error in the constraint on $f_a$ is much less than
other sources of error, in the nuclear equation of state, for
example. We conclude that our present knowledge of
the ${\Delta q}$ is adequate for the purpose of bounding $f_a$.

    We conclude that the available polarized structure function data
are in perfect agreement with QCD, once higher-order perturbative
corrections are taken into account. These are indeed essential: the
naive parton model is not adequate to describe the nucleon as viewed
through polarized lenses. The polarized structure function data
already yield an interestingly precise value of $\alpha_s$, that
future rounds of data could render highly competitive with other
determinations. The total nucleon spin fraction $\DSigma$ carried by
quarks is around 30 \%, and the strange contribution ${\Delta s
< 0}$. The dynamical mechanism that explains these findings remains
a puzzle, and the challenge to reconcile them with the \naive\
constituent quark model persists. More high-precision data would be
most welcome, in order to discriminate between the chiral soliton
and axial $U(1)$ anomaly interpretations of the data, to probe the
behaviours of $g_1^{p,n}$ at small $x$, to explore the $Q^2$-dependence
of $A_1$, and hopefully to extract higher-twist effects and $\DSigma$
from data at different values of $Q^2$.
Other experiments, for example
on elastic ${\nu N}$ scattering \cite{EK,KaplanManohar,Ahrens,nuN}
and using high-energy polarized
proton beams to measure gluon polarization directly, may also help to
disentangle the nucleon spin decomposition more completely. This is
not only a fascinating way of probing non-perturbative QCD, but also
has implications in other areas of physics, for example in searches
for Dark Matter particles, as discussed above.
{\em Floreat} nucleon spin physics!

\bigskip
\begin{flushleft}
{\bf Acknowledgements}
\end{flushleft}
We thank A. Kataev for useful conversations and for communicating
his results before publication. We also thank
S. Bethke,
M.W. Krasny and
S.~Larin
for useful discussions.
J.E. thanks the SLAC Theory Group for its
kind hospitality while this work was being completed.
The  research of M.K. was supported in part
by grant No.~90-00342 from the United States-Israel
Binational Science Foundation (BSF), Jerusalem, Israel,
and by the Basic Research Foundation administered by the
Israel Academy of Sciences and Humanities.
\bigskip
\def\etal{{\em et al.}}
\def\PL{{\em Phys. Lett.\ }}
\def\NP{{\em Nucl. Phys.\ }}
\def\PR{{\em Phys. Rev.\ }}
\def\PRL{{\em Phys. Rev. Lett.\ }}

\newpage
\begin{flushleft}
{\Large \bf Figure Captions:}\\
\end{flushleft}
\vskip 0.5cm
Fig.~1.
Values of
$\alpha_s(Q^2=2.5\ \hbox{GeV}^2)$ extracted from the E142 $\Gamma_1^n$
and E143 $\Gamma_1^p$ measurements
using the Bjorken sum rule and increasing orders
of QCD perturbation theory.
The last point    includes    the estimate
of the 4-th order perturbative coefficient
in \cite{EstimCorr}.
\hfill\break
\vskip 0.1cm
\noindent
Fig.~2.
Values of
$\alpha_s(Q^2=2.5\ \hbox{GeV}^2)$ extracted from the E142 $\Gamma_1^n$
and E143 $\Gamma_1^p$ using the Bjorken sum rule
and assuming
a given non-zero value
of the higher twist coefficient $c_{\smallHT}$ (cf.
eq.~\eqref{DeltaHT}).
Dashed lines denote error bands corresponding
to one standard deviation with respect to the central value of
$\alpha_s$.
Vertical dotted lines denote the range of $c_{\smallHT}$
given in eq.~\eqref{DeltaHT}.
\hfill\break
\vskip 0.1cm
\noindent
Fig.~3.
The value of
$\alpha_s(Q^2=2.5\ \hbox{GeV}^2)$ from the Bjorken sum rule,
using all available
perturbative QCD corrections but neglecting possible higher-twist
effects, shown
together with a compilation of world data on $\alpha_s(Q)$
as given in ref.~\cite{Bethke}.
\hfill\break
\vskip 0.1cm
\noindent
Fig.~4.
EMC, SMC and preliminary E143  $p$ data on
$\GIp$, together with $\GIp$
inferred from the E142  $^3$He and SMC D data using the Bjorken
sum rule.
The upper continuous curve shows the
{\naive}ly-suggested \cite{EJ} value holding if $\Ds = 0$,
together with an error band plotted with dotted curves.
The lower curve shows
the prediction for $\GIp$ that would be obtained
if $\Ds = -0.10 \pm 0.04$.
\hfill\break
\vskip 0.1cm
\noindent
Fig.~5.
The values of $\DSigma(Q^2{=}10\ \hbox{GeV}^2)$
extracted from each experiment plotted as functions of the
increasing order of QCD perturbation theory
used in obtaining  $\DSigma$ from the data.
The last point    includes    the estimate
of the 4-th order perturbative coefficient
in \cite{EstimCorr}.


\begin{thebibliography}{99}

\bibitem{BjI} J. Bjorken, \PR {\bf 148}(1966)1467.
\bibitem{BjII} J. Bjorken, \PR {\bf D1}(1970)1376.

\bibitem{oldSLACa}
SLAC-Yale E80 Collaboration, M.J. Alguard \etal, \PRL\break
{\bf 37}(1976)1261;
{\bf 41}(1978)70.

\bibitem{oldSLACb}
SLAC-Yale Collaboration, G. Baum \etal, \PRL {\bf 45}(1980)2000;

\bibitem{oldSLACc}
SLAC-Yale E130
Collaboration, G. Baum \etal, \PRL {\bf 51}(1983)1135;

\bibitem{EJ} J. Ellis and R.L. Jaffe, \PR {\bf D9}(1974)1444;
{\bf D10}(1974)1669.

\bibitem{Kodaira}
J. Kodaira \etal, \PR {\bf D20}(1979)627;
J. Kodaira \etal, \NP {\bf B159}(1979)99.

\bibitem{KodairaAnomDim}
J. Kodaira, \NP {\bf B165}(1979)129.

\bibitem{EMCPL} The EMC Collaboration, J. Ashman \etal,
\PL {\bf B206}(1988)364.

\bibitem{EMCNP} The EMC Collaboration, J. Ashman \etal,
\NP {\bf B328}(1989)1.

\bibitem{SmallSpinI}{J. Ellis, R. Flores and S. Ritz
\PL {\bf 198B}(1987)393.}

\bibitem{SmallSpinII}{M. Gl\"uck and E.~Reya, Dortmund report,
 DO-TH-87/14, Aug. 1987.}

\bibitem{BEK} J. Ellis, S.J. Brodsky
 and M. Karliner, \PL {\bf B206}(1988)309.

\bibitem{DeltagI}
A.V. Efremov and O.V. Teryaev, Dubna report,
JIN-E2-88-287, 1988.

\bibitem{DeltagII}
G~ Altarelli and G. Ross,
\PL {\bf B212}(1988)391.

\bibitem{DeltagIII}
R.~D.~Carlitz, J.D.~Collins and A.H.~Mueller,
\PL  {\bf B214}(1988)229;
see also
R.L.~Jaffe and A.~Manohar,
\NP {\bf B337}(1990)509;
G.T. Bodwin, J-W. Qiu,
\PR {\bf D41}(1990)2755.

\bibitem{BJviolation}
S. Forte,
\PL {\bf B309}(1993)174;
N.I. Kochelev,
{\em Vacuum QCD and new information on nucleon structure functions},
hep-ph/9307246;
I. Halperin,
{\em Veneziano ghost versus isospin breaking},
hep-ph/9312261, to appear in {\em Phys. Rev. D}.

\bibitem{SMCd}
The SMC Collaboration, B. Adeva \etal,
\PL {\bf B302}(1993)533.

\bibitem{E142}
The E142 Collaboration, P.L. Anthony \etal,
\PRL {\bf 71}(1993)959.

\bibitem{EKBjSR}
J. Ellis and M. Karliner,
\PL {\bf B313}(1993)131 and
{\em Spin Structure Functions};
Plenary talk at the 1993 PANIC XIII
Conference,
CERN-TH.7022/93, TAUP 2094-93, hep-ph/9310272.

\bibitem{BJcorr}
S.A. Larin, F.V. Tkachev and J.A.M. Vermaseren,
{\em Phys. Rev. Lett. }{\bf 66}(1991)862;
S.A. Larin and J.A.M. Vermaseren,
{\em Phys. Lett. }{\bf B259}(1991)345.

\bibitem{HigherTwist}
E.V. Shuryak and A.I. Vainshtein \NP {\bf B201}(1982)141.

\bibitem{BBK}
I.I. Balitsky, V.M. Braun and A.V. Kolesnichenko,
\PL {\bf B242}(1990)245;
erratum:
{\em ibid}, {\bf B318}(1993)648.

\bibitem{SMCp}
SMC Collaboration, D. Adams \etal,
\PL {\bf B329}(1994)399.

\bibitem{E143}
E143 Collaboration,
R. Arnold \etal, preliminary results presented at
Conference on the Intersections
of Particle and Nuclear Physics St. Petersburg, Florida,
June 1994.

\bibitem{Larin}
S.A. Larin, {\em The next to leading QCD approximation to the
Ellis-Jaffe sum rule}, CERN TH.7208/94, hep-ph/9403383.

\bibitem{EstimCorr}
A. L. Kataev and V.~Starshenko,
{\em Estimates of the $\OO (\alpha_s^4)$
corrections to $\sigma_{tot} (e^+ e^ \rightarrow \hbox{hadrons}$,
$\Gamma (\tau \rightarrow \nu_{\tau} + \hbox{hadrons})$
and deep inelastic scattering sum rules,}
CERN-TH-7198-94,
hep-ph/9405294.

\bibitem{KataevSinglet}
A. L. Kataev, private communication and
CERN TH.7333/94.

\bibitem{FoverD}
S.Y. Hsueh \etal, \PR {\bf D38}(1988)2056.

\bibitem{GLLSsr}
J. Chyla and A. L. Kataev,
{\em Phys.Lett.} {\bf B297}(1992)385;
A.~L.~Kataev and A.V.~Sidorov,
{\em The Gross Llewellyn-Smith sum rule: theory vs experiment},
CERN-TH-7235-94, hep-ph/9405254

\bibitem{alphasALEPH}
ALEPH Collaboration, D. Buskulic \etal,
\PL {\bf B307}(1993)209.

\bibitem{Petratos}
M. Petratos reporting on behalf of E142 Coll.,
Proceedings Moriond 1994 and
Kent State preprint KSUCNR-10-94.

\bibitem{NMC}
NMC Collaboration, P. Amaudruz \etal,
\PL {\bf B295}(1992)159.

\bibitem{Whitlow}
L.W. Whitlow \etal,
\PL {\bf B250}(1990)193.

\bibitem{ZvN}
E.B. Zijlstra and W.L. van Neerven,
\NP {\bf B417}(1994)61.

\bibitem{ANR}
G. Altarelli, P. Nason and G. Ridolfi,
\PL {\bf B320}(1994)152, erratum -- {\em ibid}., {\bf B325}(1994)538.

\bibitem{BBS}
S. J. Brodsky, M. Burkardt, I. Schmidt,
{\em Perturbative QCD constraints on the shape of polarized
quark and gluon distributions}, SLAC-PUB-6087, hep-ph/9401328,
submitted to {\em Nucl. Phys. B.}

\bibitem{GS}
T. Gehrmann and W.J.Stirling,
{\em Spin-dependent parton distributions from polarized structure
function  data}, hep-ph/9406212.

\bibitem{ALNR}
G. Altarelli, B. Lampe, P. Nason and G. Ridolfi,
CERN preprint TH.7254/94 (1994).

\bibitem{BassLandshoff}
S.D. Bass and P.V. Landshoff,
{\em The small x behavior of $g_1$},
CAVENDISH-HEP-94-4, hep-ph/9406350.

\bibitem{NonPertP}
P.V. Landshoff and O. Nachtmann, {\em Z. Phys.}
{\bf C35}(1987)405;
J.R. Cudell, A. Donnachie and P.V. Landshoff,
\NP {\bf B322}(1989)55;
A. Donnachie and P.V. Landshoff,
\NP {\bf B311}(1988)509;
{\em ibid}, {\bf B244}(1984)322; {\bf B267}(1985)690 and
\PL {\bf B207}(1988)319.

\bibitem{Heimann}
R.L. Heimann, \NP {\bf B64}(1973)429.

\bibitem{EK} J. Ellis and M. Karliner, \PL {\bf B213}(1988)73.

\bibitem{ThreeLoopBeta}
O.V. Tarasov, A.A. Vladimirov and A.Yu. Zharkov,
\PL {\bf 93B}(1980)429.

\bibitem{Bernreuther}
W. Bernreuther,
{\em Z. Phys.} {\bf C20}(1983)331.

\bibitem{Bethke}
S. Bethke, {\em QCD and jets at LEP},
in Proc. 22$^{nd}$ INS Symposium on Physics with
High Energy Colliders, Tokyo, March 8-10, 1994.

\bibitem{CloseRoberts}
F.E. Close and R.G. Roberts,
\PL {\bf B316}(1993)165.

\bibitem{Ji}
X. Ji and P. Unrau,
{\em $Q^2$ dependence of the proton's $G_1$ structure function
sum rule}, preprint MIT-CTP-2232, hep-ph/9308263.

\bibitem{BI}
V.D. Burkert and B.L. Ioffe,
{\em On the $Q^2$ variation of spin dependent deep inelastic electron
-- proton scattering}, preprint CEBAF-PR-92-018 (1992).

\bibitem{ESM}
B. Ehrnsperger, A. Schaefer and L. Mankiewicz,
\PL {\bf B323}(1994)439.

\bibitem{RossRoberts}
G.G. Ross and R.G. Roberts,
\PL {\bf B322}(1994)425.

\bibitem{ZZ}
Zhen-ping Li and Zhu-jun Li,
{\em On the $Q^2$ dependence of the spin structure function in the
resonance region},
hep-ph/9402308.

\bibitem{MSS}
L. Mankiewicz, A. Schaefer, E. Stein, private communication.

\bibitem{Mueller}
A.H. Mueller, \PL {\bf B308}(1993)355.

\bibitem{Krasny}{M.W. Krasny, private communication.}

\bibitem{HERMES}
M. Dueren and K. Rith,
{\em Polarized $e N$ scattering at HERA: the HERMES experiment},
in Proc. {\em Physics at HERA}, vol. I, pp. 427-445, Hamburg 1991.

\bibitem{LipkinWeak}
H.J. Lipkin, {\em How to use weak decays in analyses of data
on nucleon spin structure functions}, hep-ph/9405326.

\bibitem{CloseRobertsG1}
F.E. Close and R.G. Roberts,
{\em The Spin Dependence of Diffractive Processes and Implications
for the Small
$x$ Behaviour of $g_1$ and the Spin Content of the Nucleon},
hep-ph/9407204.

\bibitem{Ryzak}
Z. Ryzak,
\PL {\bf B217}(1989)325,
\PL {\bf B224}(1989)450.  

\bibitem{SkyrmeNow}
J. Schechter \etal, \PRL {\bf 65}(1990)2955; {\em Mod. Phys. Lett.}
{\bf A7}(1992)1;
\PR {\bf D48}(1993)339.

\bibitem{MG}{A.~Manohar and H.~Georgi,
Nucl.\ Phys.\ B{\bf 234}, 189 (1984).}

\bibitem{Kaplan}{D.~B.~Kaplan,
\PL{\bf B235}, 163 (1990);
\NP{\bf B351}, 137 (1991).}

\bibitem{Fritzsch}
H.~Fritzsch,
\PL {\bf B256}(1991)75;
\NP {\em B (Proc.Suppl.)} {\bf 23A}(1991)91
and
{\em The Internal Structure of Constituent Quarks,}
pp. 131-139 in {\em Quark Cluster Dynamics},
1992 Bad Honeff Proc.,
K. Goeke \etal, Eds.,
Springer, Heidelberg (1993).

\bibitem{EFHK}{J.~Ellis, Y.~Frishman, A.~Hanany, and M.~Karliner,
Nucl.\ Phys.\ {\bf B382}(1992)189.}

\bibitem{staticQ}
G. Gomelski, M. Karliner and S.B. Selipsky,
\PL {\bf B323}(1994)182.

\bibitem{Weise}
K. Steininger and W. Weise,
\PR {\bf D48}(1993)1433.

\bibitem{KKR}
M. Kamionkowski, L.M. Krauss and M.T. Ressell,
{\em Implications of recent nucleon spin structure measurements for
neutralino dark matter detection},
IASSNS-HEP-9414, hep-ph/9402353.

\bibitem{Mayle}
R. Mayle \etal,
{\em Phys. Lett.} {\bf B219}(1989)515.

\bibitem{KaplanManohar}
D. B. Kaplan and A. Manohar, \NP {\bf B310}(1988)527.

\bibitem{Ahrens}
L.A. Ahrens \etal, \PR {\bf D35}(1987)785.

\bibitem{nuN}
G.T. Garvey, W.C. Louis and D.H. White,
{\em Phys. Rev.} {\bf C48}(1993)761.

\end{thebibliography}
\end{document}